\documentclass[aps,prl,twocolumn,superscriptaddress]{revtex4-1}
\usepackage{graphicx,amsmath} 
\usepackage[amssymb]{SIunits} 
\usepackage{color}
\usepackage[dvipsnames]{xcolor}
\usepackage{multirow}
\usepackage{placeins}
\usepackage[overload]{empheq}

\begin{document}
\title{Dispersion of air bubbles in isotropic turbulence}
\author{Varghese Mathai${}^{*}$}
\affiliation{Physics of Fluids group, Department of Science and Technology, Max Planck Center Twente for Complex Fluid Dynamics, MESA+ Institute, and J. M. Burgers Center for Fluid Dynamics, University of Twente, P.O. Box 217, 7500 AE Enschede, The Netherlands}

\author{Sander G. Huisman}\thanks{V. Mathai and S. G. Huisman contributed equally to this work.}
\affiliation{Univ Lyon, ENS de Lyon, Univ Claude Bernard, CNRS, Laboratoire de Physique, F-69342 Lyon, France}
\affiliation{Physics of Fluids group, Department of Science and Technology, Max Planck Center Twente for Complex Fluid Dynamics, MESA+ Institute, and J. M. Burgers Center for Fluid Dynamics, University of Twente, P.O. Box 217, 7500 AE Enschede, The Netherlands}

\author{Chao Sun}
\affiliation{Center for Combustion Energy, Key Laboratory for Thermal Science and Power Engineering of Ministry of Education, Department of Energy and Power Engineering, Tsinghua University, Beijing, China}
\affiliation{Physics of Fluids group, Department of Science and Technology, Max Planck Center Twente for Complex Fluid Dynamics, MESA+ Institute, and J. M. Burgers Center for Fluid Dynamics, University of Twente, P.O. Box 217, 7500 AE Enschede, The Netherlands}

\author{Detlef Lohse}\thanks{d.lohse@utwente.nl}
\affiliation{Physics of Fluids group, Department of Science and Technology, Max Planck Center Twente for Complex Fluid Dynamics, MESA+ Institute, and J. M. Burgers Center for Fluid Dynamics, University of Twente, P.O. Box 217, 7500 AE Enschede, The Netherlands}
\affiliation{Max Planck Institute for Dynamics and Self-Organization, 37077 G\"ottingen, Germany}

\author{Micka\"el Bourgoin}\thanks{mickael.bourgoin@ens-lyon.fr}
\affiliation{Univ Lyon, ENS de Lyon, Univ Claude Bernard, CNRS, Laboratoire de Physique, F-69342 Lyon, France}

\date{\today}

\begin{abstract} Bubbles play an important role in the transport of chemicals and nutrients in many natural and industrial flows. Their dispersion is crucial to understand the mixing processes in these flows. Here we report on the dispersion of millimetric air bubbles in a homogeneous and isotropic turbulent flow with Reynolds number $\text{Re}_\lambda$ from $110$ to $310$. We find that the mean squared displacement (MSD) of the bubbles far exceeds that of fluid tracers in turbulence. The MSD shows two regimes. At short times, it grows ballistically ($\propto \tau^2$), while at larger times, it approaches the diffusive regime where MSD $\propto \tau$. Strikingly, for the bubbles, the {\it ballistic-to-diffusive} transition occurs one decade earlier than for the fluid. We reveal that both the enhanced dispersion and the early transition to the diffusive regime can be traced back to the unsteady wake-induced-motions of the bubbles. Further, the diffusion transition for bubbles is not set by the integral time scale of the turbulence (as  it is for fluid tracers and microbubbles), but instead, by a timescale of eddy-crossing of the rising bubbles. The present findings provide a Lagrangian perspective towards understanding mixing in turbulent bubbly flows.
\end{abstract}

\maketitle

Turbulent flows are ubiquitous in nature and industry. They are characterized by the presence of a wide range of length and time-scales, which enable very effective mixing. In most situations, turbulent flows contain suspended particles or bubbles -- examples are pollutants dispersed in the atmosphere, water droplets in clouds, air bubbles and plankton distributions in the oceans, and fuel sprays in engine combustion~\cite{lugt1983autorotation,la2001fluid,toschi2009lagrangian,bourgoin2014focus}. Consequently, the dispersion of suspended material by the randomly moving fluid parcels constitutes an essential feature of turbulence. 

Particle dispersion in turbulence is usually investigated by considering the mean squared displacement (MSD). As shown by Taylor in 1922~\cite{taylor1922diffusion}, for passively advected particles (or fluid), one can derive the short and long-time behaviors for the MSD:
\begin{align}
\frac{\sigma(\Delta_\tau x)^2}{\sigma(u_f)^2} = 
\begin{cases}
 \tau^2 & \text{for}~\tau \ll T_L ~~~~\text{(ballistic)}\\
 2 T_L \tau & \text{for}~\tau \gg T_L ~~~~\text{(diffusive).}
\end{cases}
\label{diff_trans}
\end{align}
Here, $\sigma(\Delta_\tau x)^2 = \left \langle \left(\Delta_\tau x - \left \langle \Delta_\tau x \right \rangle \right)^2 \right \rangle $, $\Delta_\tau x = x(t+\tau) - x(t)$, $\langle \hdots \rangle$ denotes averaging in time $t$ and ensemble averaging, $\tau$ the time lag, $\sigma(u_f)$ the standard deviation of the fluid velocity, and $T_L=\int_0^\infty C_{u_fu_f}(\tau) \textrm{d}\tau$ the Lagrangian integral timescale of the flow, with $C_{u_fu_f}$ the Lagrangian velocity autocorrelation function~\cite{taylor1922diffusion}. In eq.~(\ref{diff_trans}) the short-time ballistic regime, which is the leading order term of the Taylor-expansion of $\sigma(\Delta_\tau x)^2$ for small $\tau$, can be interpreted as a time dependent diffusion coefficient $D(\tau) = \sigma(u_f)^2 \tau$, while at larger times (and length scales) the behavior is purely diffusive with $D(\infty) = 2 \sigma(u_f)^2 T_L$ \cite{bourgoin2006role,bourgoin2015turbulent}. Importantly, $D$ exceeds the molecular diffusion coefficient by several orders of magnitude, enabling turbulence to mix and transport species much faster than can be done by molecular diffusion alone~\cite{almeras2015mixing,risso2017agitation,grossmann1984unified,grossmann1990diffusion}.

When the suspended particles are inertial, they deviate from the fluid pathlines and distribute inhomogeneously within the carrier flow~\cite{calzavarini2008quantifying}. This can lead to major differences in the particles' dispersion as compared to that of the fluid (eq.~\ref{diff_trans}). Several investigations have theoretically and numerically addressed the dispersion of inertial particles in turbulence~\cite{elghobashi1992direct,wang1993dispersion}. The advent of Lagrangian Particle Tracking has stimulated numerous experimental studies as well, on the turbulent transport of material particles~\cite{toschi2009lagrangian,bourgoin2014focus}. In particular, the dispersion of small inertial (heavy) particles has been explored in great detail. For heavy particles ($\Xi \equiv \rho/\rho_f \gg 1$), inertia can lead to enhanced MSD, while gravity induces anisotropic dispersion rates~\cite{csanady1963turbulent,maxey1987gravitational}. 

In addition to inertia and gravity, finite-size effects add to the complexity of particle dynamics in turbulent flows. Large particles filter out the small-scale fluctuations~\cite{machicoane2014large,qureshi2007turbulent, qureshi2008acceleration,volk2008acceleration,volk2011dynamics,mathai2015wake}, an effect that could be partially accounted for in the point-particle model~\cite{maxey1983equation} through the so-called Fax\'en corrections~\cite{calzavarini2009acceleration,homann2010finite}. Further, the particle's shape and even its moment of inertia can have dramatic effects on the dynamics~\cite{ern2012wake,voth2017anisotropic,mathai2018flutter,mathai2017mass}. For instance, ellipsoidal particles are known to spiral or zig-zag in flows, while disks and rods may either tumble or flutter in a flow. These can have major consequences in many applications, including sediment transport and mixing. For the long-time dispersion rate (or velocity correlation timescale), no clear consensus exists, primarily due to the experimental difficulty to access long particle trajectories. Most studies have therefore been restricted to the acceleration statistics, owing to their shorter decorrelation timescales~\cite{volk2011dynamics,volk2008acceleration,mercado2012lagrangian,mathai2016microbubbles,bec2006acceleration}.

Beyond the case of heavy particles, many practical flows contain finite-sized bubbles, typically of diameter $d_b$ in the 1--2 $\milli \meter$ range. For these, buoyancy can lead to noticeable bubble rise velocities $u_b \approx \sqrt{gd_b}$~\cite{clift1978bubbles}, where $g$ is the gravitational acceleration. In addition, their Weber number We = $\rho_f u_b^2 d_b/\gamma$ and Reynolds number $\text{Re}_b = u_b d_b/\nu$ can become large, resulting in complex interactions between the bubble and the fluid~\cite{clift1978bubbles,mougin2001path,bunner2003effect,van2008numerical,roghair2011energy,almeras2017experimental}.
Specifically for big bubbles, experimental data on the Lagrangian dynamics is scarce. One of the few existing studies is by Volk~\emph{et al.}~\cite{volk2008acceleration}, who addressed the dynamics of small, yet finite-sized bubbles ($d_b \approx 5\eta$) in an inhomogeneous von K\'arm\'an flow. Their study focused on the acceleration statistics, but did not explicitly address dispersion features. 

The present work heads to new territory through a systematic exploration of the dispersion dynamics of finite-sized millimetric bubbles in (nearly)~homogeneous isotropic turbulence (HIT). This presents several experimental challenges, as it requires a homogeneous isotropic turbulent flow seeded with a monodisperse bubble population, and the possibility to track the bubbles in 3D over timescales sufficiently large as to capture not only the small-scale dynamics, but also the large-scale dispersion and the Lagrangian velocity correlations. Features unique to the Twente Water Tunnel (TWT) facility~\cite{poorte2002experiments} such as its vertical orientation, long measurement section, flow controllability to counteract bubble rise, and an active grid that generates HIT in a large measurement volume have enabled us to achieve this.

\begin{figure}[!tbp]
	\begin{center}
		\includegraphics[width=0.49\textwidth]{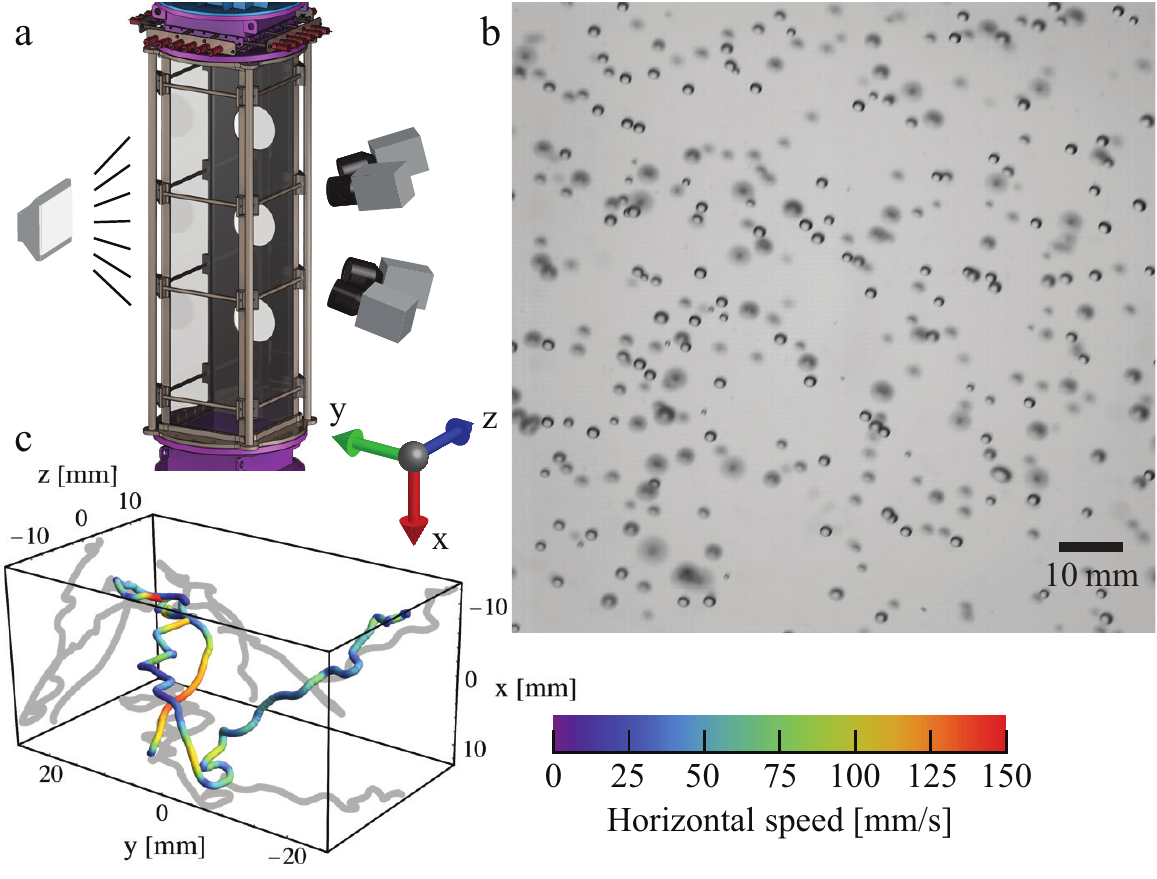}
		\caption{(a) Measurement section of the Twente Water Tunnel facility. Four high-speed cameras (Photron \texttt{1024PCI}) are mounted on the right, and the bubbles are illuminated by LED spotlights through a light diffuser. (b) Sample image by one of the cameras at $\text{Re}_{\lambda}= 150$. (c) Trajectory of a bubble (colored by horizontal speed) tracked over 595 frames for a duration of \unit{2.38}{\second} ($\approx 26 \tau_\eta$) at Re$_{\lambda} = 230$, where $\tau_\eta$ is the dissipative timescale. See also supplemental movies S1 \& S2~\cite{supvideo} of the 3D trajectories.}
		\label{fig:setup}
	\end{center}
    \vspace{-.5 cm}
\end{figure}

{\color{black}
The TWT has a measurement section of \unit{2}{\meter} length and \unit{0.4}{\meter} sides (Fig.~\ref{fig:setup}a). 	To control the turbulence, we used an active grid, driven at a random speed up to some maximum value, in a random direction, and for a random duration \cite{poorte2002experiments}. The flow was characterized by performing Constant Temperature Anemometry (CTA) measurements~(see Table~\ref{tbl:parameters}). Without bubbles, the flow may be regarded as nearly homogeneous and isotropic turbulence, with a level of anisotropy $\approx$ 5\%. With the addition of bubbles, the flow becomes locally anisotropic near the bubble wakes. However, the bubble volume fractions here~($\phi \approx 5\times 10^{-4}$) are low enough to not have any noticeable effect on the overall flow~\cite{elghobashi1994predicting}.
Bubbles were injected and selected in size by matching the terminal rise velocity with the downward flow velocity, making their vertical velocity in the lab frame small. Large (small) bubbles automatically leave the measurement section as their rise velocity is too high (low) compared to the mean flow. With the chosen downward velocity ($\left < u_f \right > \approx \unit{190}{\milli \meter \per \second}$), the naturally selected bubbles were \unit{1.78 \pm 0.16}{\milli \meter} ($5$--$6 \eta$) in diameter. Four high-speed cameras~\cite{photron} were equipped with macro lenses \cite{carlzeiss}, focusing on a joint measurement volume of \unit{\approx0.6}{\liter}. Details of the calibration model~\cite{machicoane2016improvements} can be found in the supplemental material~\cite{supvideo}. The bubbles were back-illuminated by LED lights through a diffuser plate; see an example still in Fig.~\ref{fig:setup}b. Bubbles were detected in each camera, and using particle tracking velocimetry (PTV), their 3D trajectories were obtained. Fig.~\ref{fig:setup}c shows a representative bubble trajectory in the turbulent flow. Note that $x$ ($u$) denotes the vertical, and $y$ ($v$)~\&~$z$ ($w$) the horizontal directions (bubble velocity components); see Fig.~\ref{fig:setup}. 

\begin{table}[!bp]
	\begin{tabular}{| c | c | c | c | c | c | c | c | c | c | c | c | c |}
		\hline
		$\text{Re}_\lambda$ & $\sigma (u_f)$ & \text{TI} & $\eta$ & $\tau_\eta$ & $T_L$\\
		& \unit{}{{\milli \meter \per \second}} & \% & \unit{}{\micro \meter} & \unit{}{\second} & \unit{}{\second}\\
		\hline
		110 & 15 & 7.6 &  360  & 0.13 & 1.2\\
		150 & 17 & 8.8 &  370 &  0.13& 1.6 \\
		230 & 26& 14 & 300  & 0.09 & 1.6 \\
		310 & 32 & 17 & 280 & 0.08 & 1.8 \\
		\hline
	\end{tabular}
	\caption{A summary of the turbulent flow parameters. Here, Re$_\lambda$ is the Taylor-Reynolds number, $\sigma(u_f)$ the standard deviation of the fluid velocity, TI the turbulence intensity, and $\eta$ and $\tau_\eta$ the dissipative length- and  time-scales, respectively. $T_L \approx 2\sigma(u_f)^2/(C_0\epsilon$) is the Lagrangian integral timescale of the flow, where $C_0=C_0^\infty/(1+A \ \text{Re}_\lambda^{-1.64})$ based on Sawford~\cite{sawford1991reynolds}, with $C_0^\infty = 7.0$, $A\approx 365$, and $\epsilon$ the energy dissipation rate.} 
	\label{tbl:parameters}
	\vspace{-.05 cm}
\end{table}

\begin{figure}[!tbp]
	\begin{center}
		\includegraphics{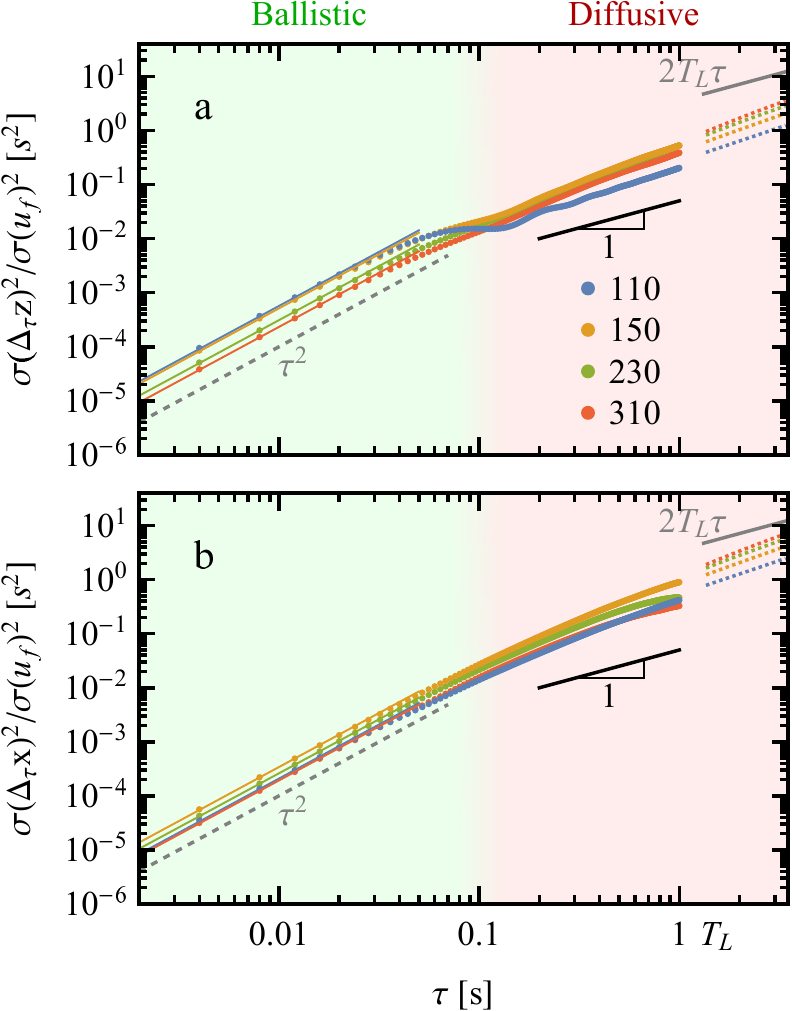}
		\caption{Mean squared displacement in the (a) horizontal and (b) vertical directions normalized by the standard deviation of fluid velocity as a function of $\tau$. The colors denote the Re$_{\lambda}$ of the experiments. For fluid tracers, the dispersion in the ballistic and diffusive regimes are given by eq.~(\ref{diff_trans}), see the gray dashed and solid  lines, respectively. The colored dotted lines in (a) \& (b) give rough predictions for the long-time-dispersion rates of the bubbles, obtained using hot-film data.}
		\label{fig:msds}
	\end{center}
    \vspace{-.4 cm}
\end{figure}

The motion of the bubbles were tracked at four different turbulence intensities TI $=\sigma(u_f)/\left <u_f \right >$, resulting in a Taylor-Reynolds Re$_\lambda$ range of $110$--$310$ (see Table.~\ref{tbl:parameters}). From the tracks we calculate the mean squared displacement (MSD), see Figs.~\ref{fig:msds}ab. The horizontal components ($z$ in Fig.~\ref{fig:msds}a) have nearly identical MSD, but the vertical component ($x$ in Fig.~\ref{fig:msds}b) is quite different. For the bubbles we observe a clear short-time behavior till $\tau \approx \unit{0.1}{\second}$, up to which the MSD grows as $\tau^2$, and with a dispersion rate well-exceeding that of the fluid (gray dashed line). Counter-intuitively, the highest horizontal dispersion for short times occurs at the lowest Re$_\lambda$~(blue data in Fig.~\ref{fig:msds}a), and decreases monotonically with increasing turbulence level. For the vertical component, the short-time dispersion rate does not show this monotonic decrease (left half of Fig.~\ref{fig:msds}b).

Beyond $\tau \approx \unit{0.1}{\second}$ the MSD appears to undergo a transition to a diffusive regime, with the local scaling exponent decreasing from $2$ to nearly $1$. At first sight this behavior seems similar to the MSD for fluid tracers~\cite{taylor1922diffusion} given by eq.~(\ref{diff_trans}). However, the {\it ballistic-to-diffusive} transition for the bubbles occurs a decade earlier in time than $T_L$.

\begin{figure}[!bp]
	\begin{center}
		\includegraphics{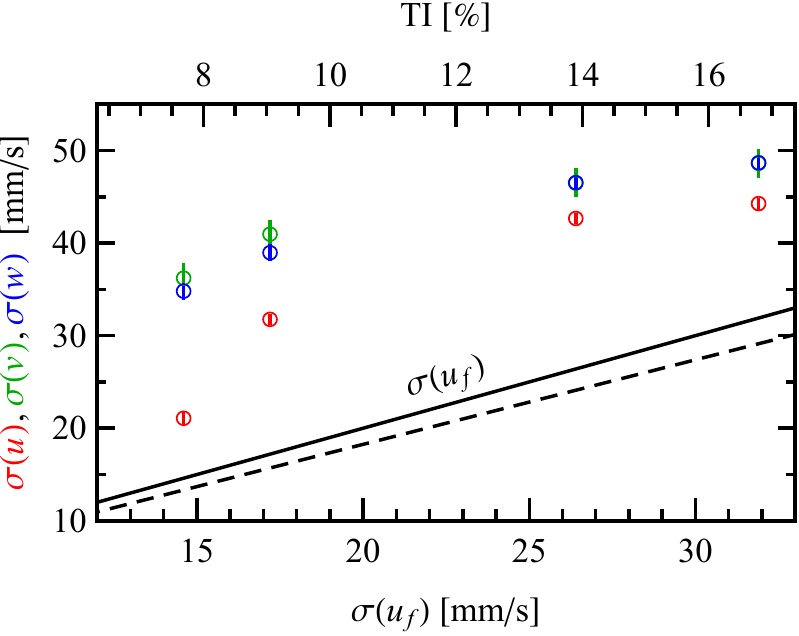}
		\caption{Standard deviation of the bubble velocities ($u, v, w$) vs the standard deviation of the flow velocity. The bubble velocity fluctuations exceed that of the fluid (solid line). The dashed line~$\sigma(u_{\text{$\mu$b}})$ is for microbubbles at Re$_\lambda = 62$~\cite{mazzitelli2004lagrangian}, obtained from fitting their data, $\sigma(u_{\text{$\mu$b}}) \approx 0.9 \sigma(u_f)$.}
		\label{fig:sigmausigmauf}
	\end{center}
    \vspace{-0.1 cm}
\end{figure}

During the transition, the lowest Re$_\lambda$ case shows oscillations before crossing over to the lowest dispersion rate (see Fig.~\ref{fig:msds}a). Thus a reversal in the behavior of the MSD occurs, wherein the lowest Re$_\lambda$ case (which had the highest ballistic dispersion) disperses the slowest. Interestingly, the diffusive regime for bubbles (right half of Fig.~\ref{fig:msds}a) lies well below the $2 T_L \tau$ found for fluid tracers in turbulence (gray line and eq.~\ref{diff_trans}).  In the vertical direction, the transition to the diffusive regime is more gradual and yields a slightly higher long-time dispersion rate as compared to the horizontal component.
In the following, we interpret these dispersion features (short-time, transitional, and long-time) in terms of the bubble dynamics and its coupling with the carrier turbulent flow.
\begin{figure}[!tbp]
	\begin{center}
		\includegraphics{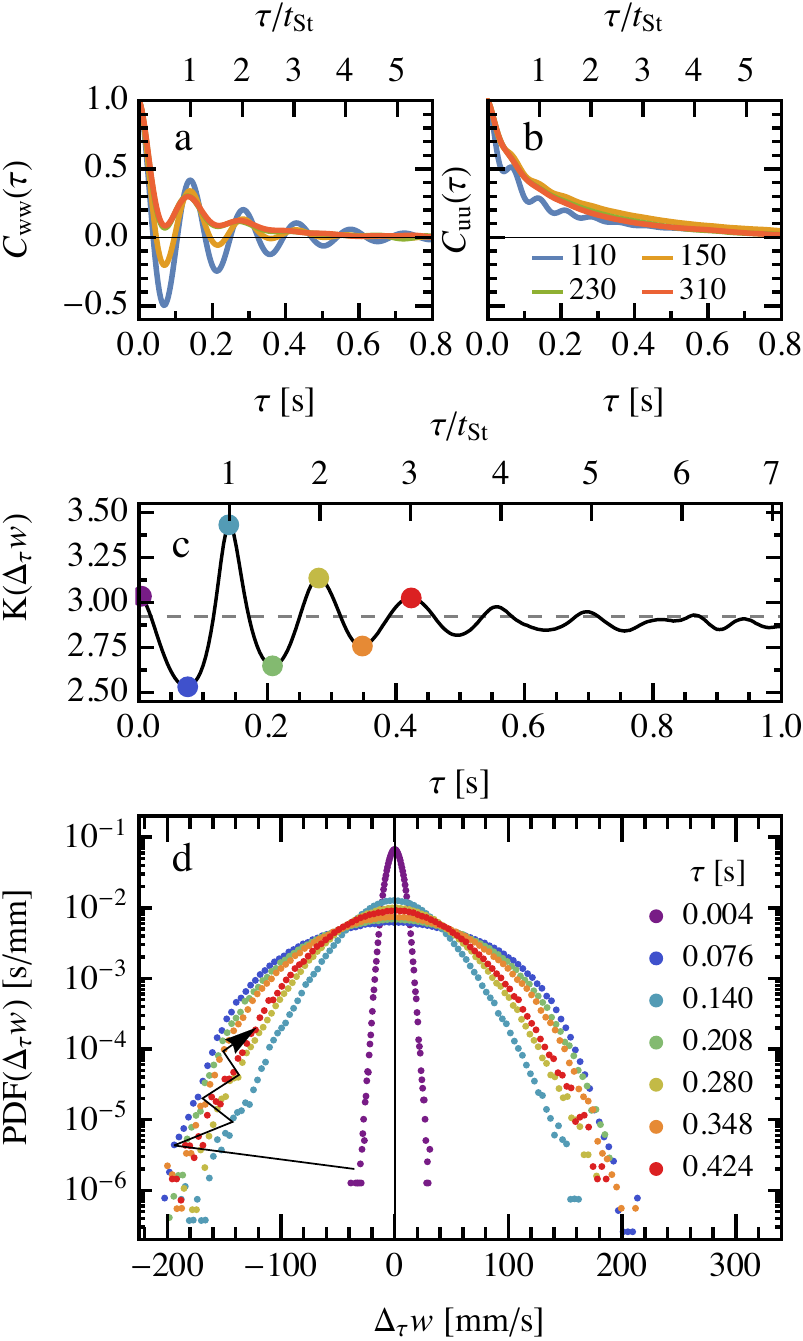}
		\caption{{Lagrangian time correlation of the (a) horizontal  and (b) vertical components of the bubble velocity. With increasing Re$_\lambda$, the oscillations dampen out. (c) Kurtosis of the horizontal velocity increment pdfs as a function of $\tau$ for the $\text{Re}_\lambda = 110$ case. For large $\tau$, the pdf converges to a sub-Gaussian kurtosis ($\text{K} \approx 2.9$), comparable to the kurtosis of the flow~\cite{almeras2017experimental}. The pdfs corresponding to the colored dots marked in (c) are given in (d). The jagged arrow connects the increment pdfs, revealing the oscillatory evolution of the pdf tails with increasing $\tau$.}}
		\label{fig:acfandmore}
	\end{center}
    \vspace{-.4 cm}
\end{figure}

The enhanced short-time dispersion can be linked to the larger velocity fluctuations of the bubbles as compared to the fluid. We fit the ballistic regime of eq.~\ref{diff_trans} to the initial ballistic trends in Fig.~\ref{fig:msds}ab. This yields the standard deviation of the bubble velocity (see Fig.~\ref{fig:sigmausigmauf}), which is much higher than that of both fluid tracers and microbubbles; all points are above the solid line for tracers---bubbles disperse faster than tracers for small times. This fast short-time-dispersion suggests a crucial mechanism responsible for the enhanced mixing in bubbly turbulent flows~\cite{risso2017agitation,almeras2015mixing}.

To explain the larger velocity fluctuations and the earlier {\it ballistic-to-diffusive} transition for the bubbles, we look into the autocorrelation function (ACF) of the bubble velocities, see Figs.~\ref{fig:acfandmore}ab. The horizontal component shows clear oscillations at a frequency $f_h = \unit{7.1} \pm$ 0.2 $\hertz$. $f_h$ shows no clear dependence on Re$_{\lambda}$ or the turbulence intensity, suggesting that the oscillations represent an intrinsic bubble frequency $f_b = u_b/d_b$. Normalizing $f_h$ by $f_b$, we obtain a Strouhal number $\text{St} \equiv \frac{f_h d_b}{u_b} \approx 0.07$, reminiscent of vortex shedding for a rising bubble~\cite{clift1978bubbles,mougin2001path,risso2017agitation}. While St is unaffected by Re$_{\lambda}$, the amplitude of the oscillations of the ACF decreases with increasing turbulence, indicating that the bubbles are decreasingly influenced by their wake-induced-motions. A similar trend is seen for the vertical component, however the oscillations are far less pronounced and occur at a frequency $f_v \approx 2f_h$ (see Fig.~\ref{fig:acfandmore}b). This clearly is characteristic of vortex-induced-oscillations, which are often twice as frequent in the stream-wise direction as in the transverse direction~\cite{mougin2001path,govardhan2005vortex}. Remnants of this are noticeable for bubbles rising in turbulent flow.

Coming back to Fig.~\ref{fig:msds}a, we can now see that the {\it ballistic-to-diffusive} transition (from $\text{MSD} \propto \tau^2$ to MSD $\propto \tau$) occurs roughly at the bubble vortex-shedding-timescale $t_\text{St} = 1/f_h \approx \unit{0.14} \second $. As discussed earlier, $t_{\text{St}}$ is set solely by $d_b$ and $u_b$, and is nearly independent of Re$_\lambda$. We note that the longest tracks are about $2 T_L$. Hence, in the high Re cases we are still within the transitional regime, with the slope gradually changing from $2$ to $1$. The vertical component shows a similar behavior (Fig.~\ref{fig:msds}b), but with a more gradual ballistic-to-diffusive transition as compared to the horizontal MSD. 

For the long-time dispersion rate, a different mechanism dominates since the buoyant bubbles drift through the turbulent eddies.
To explain this, we invoke an analogy between the velocity fluctuations experienced by the rising bubble, and those seen by a fixed-point hot-film probe in a turbulent mean flow. A rough estimate of the horizontal MSD (ignoring bubble size, inertia and path-oscillations) can be obtained from Taylor's hypothesis by considering the time-correlation function of the hot-film (CTA) signal: $\frac{\sigma(\Delta_\tau x)^2}{\sigma(u_f)^2} = 2\int_{0}^{\tau} (\tau - t)C_{u_fu_f}(t)dt$; see the colored dotted lines for $\tau>\unit{1}{\second}$ in Fig.~\ref{fig:msds}a. This Eulerian estimate works well in predicting the reduced dispersion rate of the bubbles. For the vertical component, the long-time dispersion rate is nearly twice that of the long-time horizontal MSD (see colored dotted lines in Fig.~\ref{fig:msds}b). This is consistent with the {\it continuity constraint} of incompressible flow, which causes the longitudinal integral scale to be twice its transverse counterpart for isotropic turbulence~\cite{pope2001turbulent,mathai2016microbubbles,maxey1987gravitational}. Some deviations are noticeable at higher Re$_\lambda$ because the bubble trajectories deviate from the mean vertical path at high turbulence intensities. An improved prediction might be possible if the Lagrangian time-correlation of flow velocity along the path of a rising particle were available, either from experiment~(using PIV measurements upstream of the rising bubbles) or from direct numerical simulations~(point-particles with gravity included)~\cite{parishani2015effects,calzavarini2018propelled}.

Finally, we note that the oscillatory dynamics of the bubbles affect not only the position-increment-statistics (given by the MSD), but also the statistics of Lagrangian velocity increments $\Delta_\tau w(t) = w(t+\tau) - w(t)$. This can be seen by plotting the kurtosis $K$ of the velocity increments $\Delta_\tau w(t)$ as a function of $\tau$. For fluid tracers, $K$ is known to reduce monotonically from the kurtosis of the acceleration ($K \gg 3$ for small $\tau$) to the kurtosis of the velocity~($K \approx 3$) at large $\tau$~\cite{qureshi2007turbulent}. For millimetric bubbles, we observe a non-monotonic evolution of the kurtosis~(Fig.~\ref{fig:acfandmore}c). While $K \approx 3$ at very small $\tau$, it fluctuates periodically in the range $3 \pm 0.5$ for increasing $\tau$, and finally converges to a slight sub-Gaussian value close to the kurtosis of the flow velocity ($K(u_f) \approx 2.9$~\cite{almeras2017experimental}).  For integer multiples of $t_{\text{St}}$ (top axis of Fig.~\ref{fig:acfandmore}a) the motion is positively correlated, which causes the magnitude of the increments to be smaller; resulting in distributions with higher kurtoses (see Fig.~\ref{fig:acfandmore}c). To highlight this, we calculate $\Delta_\tau w(t)$ for each trajectory and for all experiments, for the selected values of $\tau$ corresponding to minimal and maximal kurtoses from Fig.~\ref{fig:acfandmore}c. This data is then binned and rescaled to obtain the PDFs, see Fig.~\ref{fig:acfandmore}d. The jagged arrow shows the non-monotonic evolution of the tails of the velocity increment pdfs for increasing $\tau$. The oscillations are seen for all Re$_\lambda$.

To summarize, we have performed the first characterization of the dispersion dynamics of millimetric air bubbles in turbulence, a regime that has remained unexplored so far. We found that the bubbles disperse significantly faster than fluid tracers in the short-time ballistic regime: up to $6\times$ for the mean squared displacement (MSD). The transition from the ballistic regime (MSD $\propto \tau^2$) to the diffusive regime (MSD $\propto \tau$) for bubbles occurs one decade earlier than that for tracer particles, which we have linked to their oscillatory wake-driven-dynamics. In the diffusive regime, the bubbles disperse at a lower rate as compared to fluid tracers~\cite{taylor1922diffusion}, owing to their drift past the turbulent eddies. At short times, the horizontal dispersion rate dominates, while at larger times the bubbles disperse faster in the vertical direction. Our findings signal two counteracting mechanisms at play in the dispersion of bubbles in turbulence: (i) bubble-wake-oscillations, which dominate the ballistic regime and lead to enhanced dispersion rates, and (ii) the crossing-trajectories-effect, which dominates beyond the vortex-shedding-timescale of the bubbles, and contributes to a reduced long-time-dispersion. The present exploration has provided the first Lagrangian perspective towards understanding the mixing mechanisms in turbulent bubbly flows, with implications to flows in the ocean-mixing-layer and in process technology~\cite{thorpe1987bubble,pollard1990large,kantarci2005bubble,rigby2001gas}.

\begin{acknowledgments}
This work was financially supported by the STW foundation of the Netherlands, FOM, and MCEC, which are part of the Netherlands Organisation for Scientific Research (NWO), and European High-performance Infrastructures in Turbulence (EuHIT) (Grant Agreement No. 312778). We thank G.-W. Bruggert and M. Bos for technical support, and L. van Wijngaarden for discussions. We thank the referees for their input. M. Bourgoin and S. G. Huisman acknowledge financial support from the French research programs ANR-13-BS09-0009 (project LTIF). CS acknowledges the financial support from Natural Science Foundation of China under Grant No. 11672156. 
\end{acknowledgments}


\end{document}